\documentclass[11pt]{article}
\usepackage{amsmath}
\usepackage{amssymb}
\usepackage{graphicx}

\input{epsf}

\begin{document}

\begin{center}
{ \large \bf 
Quantum knitting}
\end{center}

\vspace{12pt}

\noindent 
{\large {\sl Silvano Garnerone}}\\
\noindent Dipartimento di Fisica,
Politecnico di Torino,\\
corso Duca degli Abruzzi 24, 10129 Torino (Italy)\\ 
E-mail: silvano.garnerone@polito.it \\

\noindent
{\large {\sl Annalisa Marzuoli}}\\
\noindent Dipartimento di Fisica Nucleare e Teorica,
Universit\`a degli Studi di Pavia and 
Istituto Nazionale di Fisica Nucleare, Sezione di Pavia,\\ 
via A. Bassi 6, 27100 Pavia (Italy)\\ 
E-mail: annalisa.marzuoli@pv.infn.it \\

\noindent
{\large {\sl Mario Rasetti}}\\
\noindent Dipartimento di Fisica, 
Politecnico di Torino,\\
corso Duca degli Abruzzi 24, 10129 Torino (Italy)\\ 
E-mail: mario.rasetti@polito.it \\

\begin{abstract}
We analyze the connections between the mathematical theory 
of knots and quantum physics by addressing  a number of
algorithmic questions related to both knots and braid groups.

Knots can be distinguished by means of `knot invariants', 
among which the Jones polynomial plays a prominent role, 
since it can be associated with 
observables in topological quantum field theory.

Although the problem of computing the Jones polynomial is intractable in the framework of classical
complexity theory, it has been recently recognized that a quantum computer is
capable of approximating it in an efficient way.
The quantum algorithms discussed here represent a breakthrough for quantum computation, 
since approximating the Jones polynomial is actually a `universal problem', namely
the hardest problem that a quantum computer can efficiently handle.

\end{abstract}


\vfill
\newpage

\section{Introduction}
Knots and braids, beside being fascinating mathematical objects,
are encoded in the foundations of a number of physical theories, either as 
concrete realizations of natural systems or as conceptual tools.
 The atomic model based on knot theory,
proposed in the nineteenth century, is a well known (although wrong) example of interaction
between the experimental analysis of reality, on the one hand, and
the need for classifying the observed structures in well defined mathematical categories, on the other. 
A much more recent example is provided by `knotted' configurations of DNA strands. \\
Since knots are collections of `knotted' circles  and braids are `weaving'
patterns, the term `knitting' in the title of this paper seems the more appropriate
to describe collectively these structures. Moreover, knots and braids are indeed 
closely interconnected since from a braid we can get a multi--component knot (or a `link')
by simply tying up its free endpoints, while from a knot drawn in a plane we can select  portions
which look like `over' and `under'--crossing strings ({\em cfr.} 
Fig.  \ref{knots} and Fig. \ref{closures} below).

On the conceptual side, it was in the late 1980 that knot theory was recognized to have
a deep, unexpected interaction with quantum field theory \cite{Wit}. In earlier periods of the history
of science, geometry and physics interacted very strongly at the `classical' level
(as in Einstein's General Relativity theory), but the main feature of this new,
`quantum' connection is the fact that geometry is involved in a global and not purely local
way, {\em i.e.} `topological' features do matter. To illustrate this point, consider a knot
embedded in the three--dimensional space. What really matters is
the way in which it is `knotted', while the length of the string and the bending
of the various portions of the string  itself can be changed at will (without cutting and
gluing back the endpoints). Over the years mathematicians have proposed a number of `knot
invariants' aimed to classify systematically all possible knots. As will be illustrated
in sections 2 and 3, most of these `invariants' (depending only on the topological features of the knot)
 are polynomial expressions (in one or two variables) with coefficients in the relative integers. 
It was Vaughan Jones in \cite{Jo85} who discovered the most famous polynomial invariant, 
the Jones polynomial, which  connects knots  with quantum field theory
and plays a  prominent role  in the present paper. 
In the seminal work by Edward Witten \cite{Wit}, the Jones polynomial was actually recognized to be 
associated with the vacuum expectation value 
of a `Wilson loop operator' in a particular type of three--dimensional quantum field theory
(the non--abelian Chern--Simons theory with gauge group $SU(2)$).\\
Braids  appear naturally
in this context too, since we can always `present' a knot 
as the closure of a braid. Moreover, braids enrich the purely topological nature of the theory
since the set of crossings of any braid can be endowed with a group structure.
 The Artin braid group on $n$ strands (to be defined in section 2)
encodes all topological information about `over' and `under' crossings into an algebraic setting,
opening the possibility of describing polynomial invariants of knots in terms
of `representations' of this group. Thus, the Jones polynomial can be  interpreted not only
as the `trace' of a (suitably chosen) matrix representation of the braid group \cite{Jo85}, 
but also as the  operatorial trace of an observable in a unitary quantum field theory \cite{Wit}. In section 
3 below we shall provide the (minimum)  mathematical background needed for understanding this crucial issue,
and realizing `quantum knitting'.\\

The search for new algorithmic problems and techniques 
which should improve `quantum'  with respect to classical computation is getting more and more 
challenging. Most quantum algorithms run on the standard quantum circuit model \cite{NiCh},
and are designed to solve problems which are essentially number theoretic (such as the
Shor algorithm). However, other
types of problems, typically classified in the field of enumerative combinatorics, 
and ubiquitous in many areas of mathematics and physics, 
share the feature to be `intractable' in the framework of classical information theory. 
In this latter perspective, the first section of the present paper deals with a list of
algorithmic questions in knot theory and in the theory of finitely presented
groups, focusing in particular on the braid group and its algebraic structure encoded into the
Yang--Baxter relation. These examples are  interesting not only 
for mathematicians and computer scientists, but are addressed by physicists in
the study of both exact solvable models in classical statistical mechanics \cite{Wu}
(where Yang--Baxter equation reflects the integrability properies of such models)
and quantum field theories (such as topological field theories  mentioned 
above and conformal field theories \cite{GoRuSi}).\\
In section 3 we give the essential mathematical apparatus underlying the
construction of link polynomials. This will set the stage for the last section, where 
algorithmic questions concerning the calculation of such topological invariants are raised
and discussed. Indeed, in the late 2005 Dorit Aharonov, Vaughan Jones and Zeph Landau \cite{AhJoLa} 
found an efficient quantum algorithm for approximating the Jones polynomial
(a difficult problem to be addressed in the classical context), thus giving a
fundamental improvement of previous claims on the computational complexity of such problem 
\cite{BoFrLo}. Two papers of ours \cite{GaMaRa1,GaMaRa2} provided essentially similar 
results for a wider  class of link invariants, and
the paper by Wocjan and Yard \cite{WoYa} addressed and extended the algebraic 
and information--theoretical  background used in \cite{AhJoLa}.
The quantum algorithm in question (reviewed and discussed  in section 3) 
represents a breakthrough for quantum computation
because the problem of approximating
the Jones polynomial has been recognized to be a `universal problem',
namely the hardest problem that a quantum computer can efficiently handle.
The latter remark sounds reminiscent of  `quantum knitting': we can
really knit knots and braids by means
of  `quantum' computing machines.

\section{Knots, braids and related classical \\
algorithmic problems}

A {\em knot} $K$ is defined as a continuous embedding of the circle $S^1$ (the $1$--dimensional 
sphere) into the euclidean $3$--space $\mathbb{R}^3$ or, equivalently, into the $3$--sphere
$S^3 \doteq \mathbb{R}^3 \cup \{\infty\}$. A {\em link} $L$ is the embedding of the disjoint 
union of $M$ circles, $\cup_{m=1}^{M}\,(S^1)_m$ into $\mathbb{R}^3$ or $S^3$, namely a finite 
collection of knots referred to as the components of $L$ and denoted by 
$\{L_m\}_{m=1,2,\ldots,M}$. Since each circle can be naturally endowed with an orientation, 
we can introduce naturally {\em oriented} knots (links).

Referring for simplicity to the unoriented case, two knots $K_1$ and $K_2$ are said to be 
{\em equivalent}, $K_1 \sim K_2$, if and only if they are (ambient) isotopic. An isotopy
can be thought of as a continuous deformation of the shape of, say, $\,K_2 \subset \mathbb{R}^ 3$
which makes $K_2$ identical to $K_1$ without cutting and gluing back the `closed string' 
$K_2$.

\begin{figure}[htbp]
\begin{center}
\includegraphics[height=6cm]{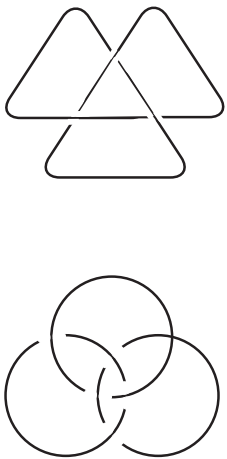}
\end{center}
\caption{Planar diagrams of the trefoil knot (top) and  Borromean link (bottom).}
\label{knots}
\end{figure}

The {\em planar diagram}, or simply the {\em diagram}, of a knot $K$ is the projection of $K$
on a plane $\mathbb{R}^2 \subset \mathbb{R}^3$, in such a way that no point belongs to
the projection of three segments, namely the singular points in the diagram are only
transverse double points. Such a projection, together with `over' and `under' information
at the crossing points --depicted in figures by breaks in the under--passing segments--
is denoted by $D(K)$; a {\em link diagram} $D(L)$ is defined similarly. In what follows we 
shall sometimes identify the symbols $K$ [$L$] with $D(K)$ [$D(L)$], although we can 
obviously associate with a same knot (link) an infinity of planar diagrams.
Examples of diagrams are depicted in Fig. \ref{knots}.

\begin{figure}[htbp]
\begin{center}
\includegraphics[height=6cm]{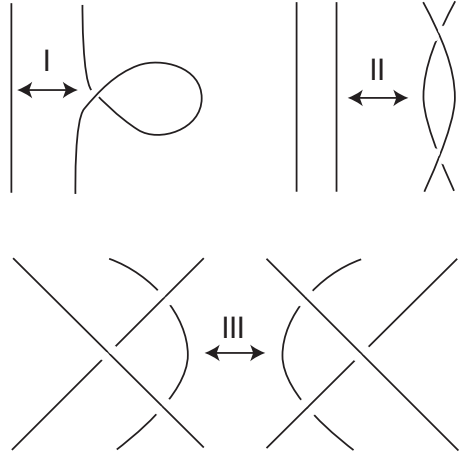}
\end{center}
\caption{The three Reidemeister moves acting on local configurations in link diagrams.}
\label{reidemeister}
\end{figure}

After these preliminary definitions, let us state the first of our question on
algorithmic complexity, actually the fundamental, still unsolved problem in knot theory.
\begin{quote}
{\em \underline{Problem 0}.  Give an effective algorithm for establishing when two knots
or links are equivalent.}
\end{quote}
The number of crossings of a knot (diagram) is clearly a good indicator of the `complexity'
of the knot. Indeed, Tait in  late 1800 initiated a program aimed to classifying sistematically 
knots in terms of the number of crossings (see \cite{Rol}, \cite{Bir, BiBr},  
\cite{Lic} for exhaustive 
accounts on knot theory and for references to both older papers and knots tables).\\ 
Since a knot with crossing number $\kappa \equiv c(K)$ can be represented by planar diagrams
with crossing numbers $c(D(K))$ for each $c(D(K))>\,\kappa$, the first issue to
be addressed is the search for procedures aimed to simplify 
as much as possible the diagrams of a knot $K$ to get a $D'(K)$ with $c(D'(K))=\kappa$,
the `minimum' crossing number.
Reidemeister's theorem (see \cite{Bir}) gives the answer to this basic question.
\begin{quote}
{\bf Equivalence of knots (Reidemeister moves).} Given any pair of planar diagrams $D,D'$ of the same
knot (or link), there exists a finite sequence of diagrams
\begin{equation}\label{Reid}
D\,=\,D_1\,\rightarrow\,D_2\,\rightarrow \dotsm \rightarrow\,D_k\,=\,D'
\end{equation}
such that any $D_{i+1}$ in the sequence is obtained from $D_i$ by means of the three
Reidemeister moves (I, II, III) depicted in Fig. \ref{reidemeister}.
\end{quote}

The procedure determined by Reidemeister's theorem applies to subsets of link diagrams
localized inside disks belonging to the plane where the diagram lives, and can be
suitably generalized to handle diagrams of oriented links.
However, notwithstanding the recursively numerable character of the implementation
of the Reidmeister moves with respect to the intractability of the notion of ambient 
isotopy, such moves cannot be formalized into effective algorithms, basically because
the above  definition is purely topological. As we shall see, transformations
on link diagrams can be consistently translated into an algebraic setting by exploiting
their deep connection with braid groups. In the new setting the `moves'
on link diagrams will be reformulated in terms of algebraic operations,
see Markov's theorem below.

Before addressing this issue, which will generate several interesting algorithmic
problems, let us point out that in geometry (as well as in physical
field theories whose dynamical variables have a geometric nature) 
an important role is played by `topological invariants'. To illustrate this point
and clarify its connection with the systematic classification of geometric structures,
consider the case of surfaces. In particular, any smooth, closed and oriented
surface $\mathcal{S}$ has a well defined topological structure, completely determined
by its Euler number $\chi(\mathcal{S})=2-2g$, where $g$ is the number of `handles'
of $\mathcal{S}$. Thus we have the $2$--sphere $S^2$ for $g=0$ (no handles), the
$2$--torus (the surface of a doughnut) for $g=1$, and more complicated, many--handles 
surfaces as $g$ grows. The existence of such a `complete' invariant in $2$--dimensional
geometry is quite an exception: higher dimensional geometrical objects (smooth 
$D$--manifolds) do not share this feature, and the associated classification problems are
still open (except for some restricted classes).

Coming back to knot theory, we define a {\em link invariant} through a map
\begin{equation}\label{linkinv}
L\;\;\longrightarrow\;\;f(L),
\end{equation}
where the quantity $f(L)$ depends only on the type of the link, namely takes
different values on inequivalent links. Switching to link diagrams, 
we keep on using the same  notation as in (\ref{linkinv}), but now it is 
sufficient to verify that $f(L)$ ($\equiv f(D(L))$) does not change
under applications of the Reidemeister moves I, II, III (Fig. \ref{reidemeister}).

We have already met a numerical invariant, namely the (minimum) crossing
number $\kappa$. It is a natural number which takes the value  $0$
for the trivial knot represented as an unknotted  circle.  
Other invariants taking values in $\mathbb{Z}$ (relative intergers) can be defined
for oriented link diagrams, where each crossing is marked by $\pm 1$ according
to some fixed convention. For instance, the {\em writhe} $w(D(L))$
of a diagram $D$ of an oriented link $L$ is the summation of the signs of the 
crossings of $D$, namely 
\begin{equation}\label{writhe}
w(D(L))\,=\,\sum_{p}\,\epsilon_p,
\end{equation}
where the sum runs over the crossing points $\{p\}$ and $\epsilon_p =+1$ if the (directed)
knot path shows an overpass at the crossing point $p$, $\epsilon_p =-1$ for an underpass.
Note however that both the crossing number and the writhe do change under Reidmeister move
of type I, but are invariant under the moves II and III: this property defines
a restricted kind of isotopy, commonly referred to as {\em regular
isotopy}. The concept of regular isotopy is very useful because, by eliminating
the move I, we do not really lose any information about the topology of the link.
Moreover, the evaluation of crossing numbers and writhes can be carried out
efficiently by a simple inspection of the diagrams.

Over the years, mathematicians have provided  a number of knot invariants, by resorting 
to topological, combinatorial and algebraic methods. Nevertheless, we do not have yet a complete 
invariant (nor a complete set of invariants) able to characterize the topological type of 
each knot and to distinguish among all possible inequivalent knots, 
{\em cfr.} Problem 0.\\
As a matter of fact, the most effective invariants have an algebraic origin, 
being closely related to the braid group and its representation theory.
As we shall see below, it is straightforward to obtain a knot (link) out
of a `braid'. The inverse process is governed by Alexander's theorem (see \cite{BiBr}, section 2).
\begin{quote}
{\bf Braids from links.} Every knot or link in $S^3=\mathbb{R}^3 \cup \{\infty\}$
can be represented as a closed braid, although not in a unique way.
\end{quote}
The {\em Artin braid group} $\mathbf{B}_n$, whose elements are (open) braids $\beta$,
is a finitely presented group on $n$ `standard' generators
$\{\sigma_1,\sigma_2,\ldots,
\sigma_{n-1}\}$ plus the identity element $e$, which satisfy the relations

\begin{align}\label {algYB1}
\sigma_i\,\sigma_j\,= &\,\sigma_j\,\sigma_i \;\;\;\;(i,j=1,2,\ldots,n-1)\;\;
\text{if}\;\,\,|i-j| > 1 \nonumber \\
\sigma_i\,\sigma_{i+1}\,\sigma_i\,= &\,\sigma_{i+1}\,\sigma_{i}\,\sigma_{i+1}
 \;\;\;(\,i=1,2,\ldots,n-2).
\end{align}

\begin{figure}[htbp]
\begin{center}
\includegraphics[height=6cm]{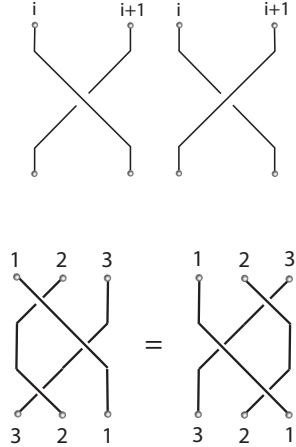}
\end{center}
\caption{The generator $\sigma_i$ and its inverse $\sigma_i^{-1}$ (top). The algebraic Yang--Baxter
equation (bottom).}
\label{braids}
\end{figure}

\noindent This group acts naturally on topological sets of $n$ disjoint strands 
with fixed endpoints --
running downward and labeled from left to right --
in the sense that each generator $\sigma_i$ corresponds to a crossing of two contiguous 
strands labeled by $i$  and $(i+1)$, respectively: if $\sigma_i$ stands for the crossing
of the $i$--th strand over the $(i+1)$--th one,
then $\sigma_i^{-1}$ represents the inverse operation with 
$\sigma_i\,\sigma_i^{-1}$ $=\sigma_i^{-1}\sigma_i =\,e$, see Fig. \ref{braids}.
An element of the braid group can be thought of as a `word', such as for instance
$\beta= \sigma_3^{-1}\sigma_2$ $\sigma_3^{-1}\sigma_2 \, 
\sigma_1^{3}$ $\sigma_2^{-1}\sigma_1\sigma_2^{-2}$ $\in \mathbf{B}_4$; the length 
$|\beta|$
of the word $\beta$ is the number of  its letters, where by a `letter' we mean one of the
generators or its inverse element.\\
By a slight change of notation, denote by $\mathsf{R}_{ij}$ 
the over--crossing operation acting on
 two strands the endpoints of which are labelled by $i$ and $j$. 
Then the second relation in
\eqref{algYB1} can be recasted into the form

\begin{equation}\label{algYB2}
\mathsf{R}_{12}\,\mathsf{R}_{13}\,\mathsf{R}_{23}\;=\;\mathsf{R}_{23}\,
\mathsf{R}_{13}\,\mathsf{R}_{12}
\end{equation}

\noindent and represented pictorially as in Fig. \ref{braids}, where operations are ordered downward.
Note that this picture can be viewed as a portion of an $n$--strands configuration  (and thus
$\{1,2,3\}$ may actually represent labels attached to any triad of contiguous strands) 
since the first
relation in \eqref{algYB1} ensures that other types of crossings cannot happen at all.
The relation \eqref{algYB2} is referred to as the 
algebraic {\em Yang--Baxter relation} and
characterizes the  structure of
solvable models in statistical mechanics 
\cite{Wu,GoRuSi}. 

\begin{figure}[htbp]
\begin{center}
\includegraphics[height=6cm]{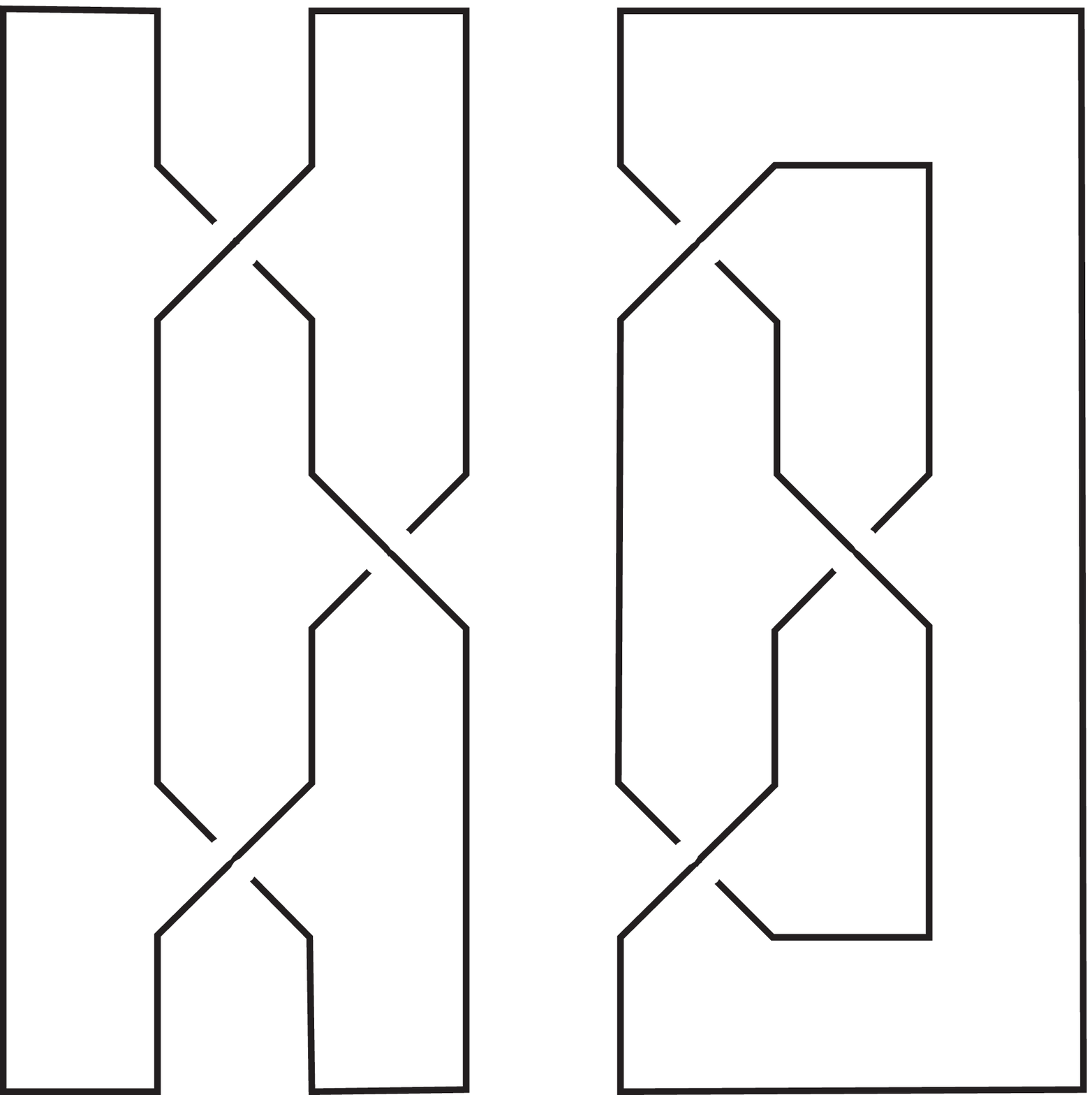}
\end{center}
\caption{The two types of closures of a braid, namely the plat closure (left)
and the standard closure (right), both representing the same trefoil knot.}
\label{closures}
\end{figure}

As mentioned before, it is straightforward to get a link out of a braid: we have simply
to `close up' the ends of an open braid $\beta$ to get a {\em closed braid} 
$\hat{\beta}$ that reproduces the diagram of some link $L$. Formally
\begin{equation}\label{beta}
\beta\;\xrightarrow{\text{closure}}\;\hat{\beta}\;\longleftrightarrow\;L\,.
\end{equation}
Notice however that this operation can be performed in two ways, denoted by
$\hat{\beta}^{\,\text{st}}$ (the standard closure) and
$\hat{\beta}^{\,\text{pl}}$ (the plat closure), respectively. In Fig. \ref{closures}
the two admissible closures of a same open braid are shown, and moreover
the content of Alexander's theorem is made manifest, since both these closed braids can be
seen as deformations (isotopy transformations) 
of the planar diagram of the trefoil knot depicted in Fig. \ref{knots}.

As already pointed out, Alexander's theorem does not 
establish a one--to--one correspondence 
between links and braids. For instance, given a closed braid
$\hat{\beta}\,=\,L$ with $\beta \in \mathbf{B}_n$, any other braid obtained from 
$\beta$ by {\em conjugation}, namely $\beta'\,=\,\alpha \beta \alpha^{-1}$
(for some $\alpha \in \mathbf{B}_n$) has a closure $\hat{\beta'}$  which reproduces 
the same link $L$. Thus the following question can be naturally rised.\\
\begin{quote}
{\em \underline{Problem 1A}. Is it always possible to transform efficiently a given 
knot or link  into a closed braid? }
\end{quote}
The answer is affirmative, since there exists a classical algorithm which
performs the reduction in a number of steps which is bounded from above by
a polynomial function of the braid index (\cite{BiBr}, section 2 and original references therein),
where the {\em braid index} of a braid or closed braid is simply the number of its strands.

Taken for granted the above result about the efficiency of the reduction
of any link diagram to a closed braid, we can now exploit the algebraic properties 
of braid groups. For what concerns in particular the issue of equivalence,
Reidemeister's theorem can be recasted
into Markov's theorem. The following statement of this theorem refers to the
case of open braids, which captures the crucial features of the construction, while
the version involving closed braids can be found in \cite{BiBr}, section 2.
\begin{quote}
{\bf Equivalence of braids (Markov moves).} Two braids are equivalent if they differ by a finite sequence of 
Markov `moves' of the following two types, together with their inverse moves:\\
{\bf i)} change a braid $\beta \in \mathbf{B}_n$ to a conjugate element
in the same group, $\beta\, \rightarrow \alpha \beta \alpha^{-1}$, with $\alpha
\in \mathbf{B}_n$;\\
{\bf ii)} change $\beta \in \mathbf{B}_n$ to $\mathsf{i}_n\,(\beta)\sigma_n^{\pm 1}$,
where $\mathsf{i}_n :\mathbf{B}_n \rightarrow \mathbf{B}_{n+1}$ is the natural inclusion
obtained by disregarding the $(n+1)$th strand and $\sigma_n, \sigma_n^{\,-1}$ 
$\in \mathbf{B}_n$.
\end{quote} 
The next question arises in connection with the search for the most 
`economical' representation of a knot diagram as a closed braid.
The {\em minimum braid index} of a link $L$ is the minimum number $n$
for which there exists a braid $\beta \in \mathbf{B}_n$ whose closure $\hat{\beta}$
represents $L$.
\begin{quote}
{\em \underline{Problem 1B}. Does there exist an (efficient) algorithm
to select, among all the diagrams of a given link $L$, the diagram with
the minimum braid index?}
\end{quote}
At present no explicit  algorithm for addressing this  problem is known, 
so that its computational 
complexity class cannot be even evaluated (see \cite{BiBr}, section 2 and 4 for more details).\\

Coming to algorithmic problems characterized in purely algebraic terms,
let us recall that braid groups belong to the class of 
{\em finitely presented groups}. Such groups are defined by means
of a finite sets of {\em generators} together with {\em relations}
among the generators  and can have a finite order --as for
the classical point groups of cristallography-- or not --as happens 
for the braid group-- (see \cite{MaKaSo} and older references therein). 
It was Max Dehn who stated the three 
`fundamental problems' (listed below) concerning a group $G$ 
presented in terms of generators, denoted by
$a,b,c,\ldots$, and relations $P,Q,R,\ldots$, namely
\begin{equation}\label{Gpres}
G\;\doteq\;\langle\, a,b,c,\ldots\,;P,Q,R,\ldots \rangle\;.
\end{equation} 
Any element of $G$ can be written (in multiplicative notation)
as a {\em word} $W$ in the alphabeth given by the generators and
their inverse elements, as already done for the braid group.
Note that the relations $P,Q,R,\ldots$ in (\ref{Gpres}) represent 
the minimal set of words in the generators, equivalent to the identity element $e$
(`minimal' meaning that any other word equivalent to the identity
can be reduced --by the use of the relations in the set--
to the union of words in the same set).\\
A simple example of relation for a group $G$ on two generators, $a,b$, is
$ab=ba$ or, equivalently, $P=ab(ab)^{-1}=e$.
Of course this means that the group is commutative, 
and looking back at the presentation of
the braid group given in (\ref{algYB1}) we see that the generators $\sigma_i$
and $\sigma_j$ do commute when $|i-j| >1$, but 
$\mathbf{B}_n$ is not a commutative group since 
$\sigma_i \sigma_{i+1}\,\neq\,\sigma_{i+1} \sigma_i$.\\
Given the presentation (\ref{Gpres}) of the group $G$, Dehn's problems
are formulated as follows.
\begin{quote}
{\em \underline{2A. The word problem.} For an arbitrary word $W$ in the generators,
decide whether or not $W$ defines the identity element in $G$.\\
 Equivalently: given two words $W$, $W'$, decide whether $W=W'$.}
\end{quote}
\begin{quote}
{\em \underline{2B. The conjugacy problem.} For two arbitrary words $W_1$ and $W_2$
in the generators, decide whether or not they are conjugate to each other.
In a sharper form: find explicitly an element $W'$ for which
$W_2=W'W_1(W')^{-1}$.}
\end{quote}
\begin{quote}
{\em \underline{2C. The isomorphism problem.} For an arbitrary group $G'$ 
defined by another presentation $G'\doteq
\langle a',b',c',\ldots;P',Q',R',\ldots \rangle$, decide whether or not
$G$ and $G'$ are isomorphic.}
\end{quote}
Except for the second issue in Problem 2B, we are in the presence of 
{\em decision problems}, namely problems that can be addressed
by means of classical, deterministic
algorithms (running on a Turing machine) designed to answer
`yes' or `no' to 
each of the above questions. Recall also that the {\em time 
complexity function} $\mathfrak{f}_{\mathcal{A}}$ of an algorithm 
$\mathcal{A}$ is defined in terms of
the size (length) $\mathfrak{s}$ of the input. In particular,
an algorithm associated with a decision problem belongs
to the complexity class $\mathbf{P}$ if $\;\mathfrak{f}_{\mathcal{A}}$ 
$(\mathfrak{s})$ is a polynomial function of $\mathfrak{s}$ and to the class
$\mathbf{NP}$ is any guess on the answer can be checked in polynomial time.
Algorithmic problems endowed with complexity functions of exponential type
are to be considered as intractable in the framework of classical information theory.

For what concerns the list above in the case of a generic group $G$, 
note first of all that a solution of the complete
conjugacy problem 2B (belonging to a certain complexity class) would 
imply a solution to Problem 2A (in the same class) since it would be sufficient
to set $W'=e$  in the expression $W_2=W'W_1(W')^{-1}$. It also clear that the most
difficult problem is the last one, which requires 
a `global' inspection of the algebraic structures (generators + relations) 
of the groups under examination. As for the braid group, we leave aside 
this problem and refer the reader to \cite{BiBr} (section 1) for the definition 
of a presentation in terms of generators and relations alternative to
the standard ones collected in (\ref{algYB1}).

The known results about the word and the conjugacy problems
for the braid group $\mathbf{B}_n$ are briefly summarized 
below (see \cite{BiBr}, section 5 and original references therein).
\begin{quote}
{\em 2A.} The solution to the word problem is polynomial, with a complexity
function of the order 
\begin{equation}\label{wordpr}
\mathcal{O}(\,|\beta|^2\,n\,\ln \,n), 
\end{equation}
where
$|\beta|$ is the length of the initial representative of the braid $\beta$
and $n$ is the braid index.  
\end{quote}
Surprisingly enough, the following problem, apparently very closely
related to the word problem, turns out to belong to the class
of $\mathbf{NP}$--complete problems (recall that the 
$\mathbf{N}$on--deterministic $\mathbf{P}$olynomial class
contains decision problems for which a guess solution can be checked in
polynomial time; a particular $\mathbf{NP}$ problem is `complete' 
if every other problem in the class can be polynomially reduced to it).
\begin{quote}
{\em \underline{Problem 2A'}. Given a word $\beta$ in the standard generators 
$\sigma_1,\sigma_2,$ $\ldots,\sigma_{n-1}$
and their inverses, determine whether there is a shorter word $\beta'$ 
which represents the same element in $\mathbf{B}_n$.              }
\end{quote}
Finally
\begin{quote}
{\em 2B.} The best known algorithm for the conjugacy problem is exponential in
both $|\beta|$  and $n$.
\end{quote}
It is worth noticing that the  difficulty of solving the
conjugacy problem in braid groups 
has been exploited for the construction 
of a public--key cryptosystem \cite{AnAnGo}.

\section{Polynomial invariants of knots via braids}

Invariants of knots (links) of polynomial type arise (or can be reformulated)
by resorting to {\em representations} of the braid group.
Generally speaking, in order to represent the finitely presented group 
$\mathbf{B}_n$ defined in
(\ref{algYB1}),
we need an `algebra' structure $\mathsf{A}$, namely a vector space 
over some field (or ring) $\Lambda$, endowed with a multiplication
satisfying associative and distributive laws. The algebra must have a
unit with respect to multiplication and for our purposes must be also finitely
generated, namely its elements can be decomposed in terms of some 
finite `basis' set, the number of elements of which equals the braid index $n$.\\
The reason for considering an algebra should become clear if we recognize, 
on the one hand, that we can multiply braids $\in \mathbf{B}_n$ by simply
composing their diagrams: given $\beta_1$ and $\beta_2$ $\in \mathbf{B}_n$
we get the product $\beta_1\,\beta_2$ by placing the braid $\beta_1$
above $\beta_2$ and gluing the bottom free ends of $\beta_1$ with the
top ends of $\beta_2$ (this opearation was implicitly assumed in 
(\ref{algYB1}) and (\ref{algYB2}), see also Fig. \ref{braids}). 
On the other hand, the operation associated with 
`addition' of braids can be defined in terms of formal combinations
of the type $a\beta_1 +b\beta_2$, for any $\beta_1,\beta_2$ $\in \mathbf{B}_n$
and $a,b \in \Lambda$ (the field of scalars associated with the algebra $\mathsf{A}$).\\
With these premises, a {\em representation of} $\mathbf{B}_n$ {\em inside the algebra}
$\mathsf{A}$ is a map
\begin{equation}\label{repr1}
\rho_{\,\mathsf{A}}\;\;: \mathbf{B}_n\,\;\longrightarrow\,\mathsf{A}
\end{equation}
which satisfies
\begin{equation}\label{omom}
\rho_{\,\mathsf{A}}(\beta_1\,\beta_2)\,=\,\rho_{\,\mathsf{A}}(\beta_1)
\rho_{\,\mathsf{A}}(\beta_2)\;\;\;\;\forall\, \beta_1\,,\beta_2 \in \mathbf{B}_n,
\end{equation}
namely $\rho_{\,\mathsf{A}}$ is a group homomorphism from $\mathbf{B}_n$
to the multiplicative group $\mathsf{G}\,\subset \mathsf{A}$ of the invertible
elements of $\mathsf{A}$ (in particular: $\rho_{\,\mathsf{A}}(e)$
$=1$, where $e$ is the identity element of
$\mathbf{B}_n$ and $1$ denotes the unit of $\mathsf{A}$; 
$\rho_{\,\mathsf{A}}(\beta^{-1})$
$=[\rho_{\,\mathsf{A}}(\beta)]^{-1}$, $\forall \beta$).
By using the standard generators of $\mathbf{B}_n$ defined in
(\ref{algYB1}), it suffices to define the map (\ref{repr1}) on the 
elements of $\{\sigma_i\}$
\begin{equation}\label{reprsigma}
\rho_{\,\mathsf{A}}(\sigma_i)\,\doteq\, g_i\,\in \mathsf{G}\,\subset\,\mathsf{A}\,,
\;\;(i=1,2,\ldots n-1),
\end{equation}
and extend linearly its action on products and sums of braids.
Any pair of contiguous elements $g_i$ and $g_{i+1}$ must satisfy the 
{\em Yang--Baxter equation associated with the representation}
$\rho_{\,\mathsf{A}}$, namely
\begin{equation}\label{reprYB}
g_i\,g_{i+1}\,g_i\,=\,g_{i+1}\,g_i\,g_{i+1}
\end{equation}
while $g_i g_j\,=\,g_j g_i$ for $|i-j|>1$.

Once defined the representation $\rho_{\,\mathsf{A}}$ we may also introduce
associated {\em matrix representations} of some fixed dimension $N$ by representing
$\mathsf{A}$ over the algebra of $(N \times N)$ matrices with entries
in the field $\Lambda$
\begin{equation}\label{matrepr}
\mathsf{A}\;\longrightarrow\;\mathsf{M} (\Lambda\,,N).
\end{equation}
If we restrict the domain of the above map to the group $\mathsf{G} \subset \mathsf{A}$
of invertible elements, the assignment (\ref{matrepr}) can be rephrased as the choice
an $N$--dimensional vector space $V$ over $\Lambda$, and thus we have the
natural isomorphism
\begin{equation}\label{matrisom}
\mathsf{M} (\Lambda\,,N)\;\cong\;\mathsf{GL}_{\,\Lambda}\,(V,\,N),
\end{equation}
where $\mathsf{GL}_{\,\Lambda}\,(V,\,N)$ is the general linear group of non--singular, 
$\Lambda$--linear maps $V \rightarrow V$.

Loosely speaking, if we associate with a braid $\beta \in \mathbf{B}_n$
a matrix $M(\beta)$ obtained by means of a representation (\ref{matrepr})
of dimension $N=n$, then
$\beta$ can be characterized by a scalar, namely the trace of $M(\beta)$
(the `character' of the representation in the group--theoretic 
language). 
Such traces are candidates to be interpreted as invariants
of links presented as closed braids, {\em cfr.} 
(\ref{beta})
in the previous section.\\
A {\em trace function over the algebra} $\mathsf{A}$ is formally defined as a linear
function over $\mathsf{A}$ and, by extension, over a matrix
representation algebra  (\ref{matrepr}) 
\begin{equation}\label{trace1}
\mathsf{A}\;\longrightarrow\;\mathsf{M} (\Lambda\,,N)\;\xrightarrow{\text{Tr}}\,
\Lambda
\end{equation}
satisfying the property
\begin{equation}\label{trace2}
\text{Tr}\,(\,M(\beta)\, M'(\beta')\,)\,=\,\text{Tr}\,(\,M' (\beta')\, M(\beta)\,).
\end{equation}
for any $M(\beta)\,, M'(\beta')$ which are the images under  $\rho_{\,\mathsf{A}}$
of two braids $\beta, \beta'$ $\in \mathbf{B}_n$. It can be shown that 
$\text{Tr}(M(\beta))$ is a link invariant since it does not change under Markov move
of type {\bf i)} (defined in section 1), namely 
\begin{equation}\label{trace3}
\text{Tr}\,(M(\beta))\,=\,\text{Tr}\,(M' (\beta' ))\;\;\;\text{if}\;\beta\;\text{and}\;
\beta' \text{are conjugate}.
\end{equation}
In other words, link invariants arising from these `Markov traces' are 
regular isotopy invariants (we leave aside the issue of the invariance under 
Markov move of type {\bf ii)}). 

The general algebraic setting outlined above is the framework
underlying the constructions of both the Jones \cite{Jo85} and the HOMFLY \cite{HOMFLY}
link polynomials. In particular:
\begin{itemize}
\item the {\em Jones polynomial} of a link $L$, $\mathsf{J}(L;t)$, is the Markov trace
of the representation of $\mathbf{B}_n$ inside the Temperley--Lieb algebra
$TL_n(t)$ (\cite{BiBr}, section 2, \cite{Jo05}). It is a Laurent polynomial in one 
formal variable $t$ with coefficients
in $\mathbb{Z}$, namely it takes values in the ring $\Lambda \equiv
\mathbb{Z}[t,t^{-1}]$;
\item the {\em HOMFLY polynomial} $P(L;t,z)$ is obtained as a one--parameter family
of Markov traces (parametrized by $z$) of the representation of 
$\mathbf{B}_n$ inside the Hecke algebra $H_n(t)$ (\cite{BiBr},  section 4, \cite{Lic}).
It is a Laurent polynomial in two formal variables with coefficients 
in $\mathbb{Z}$, namely $\Lambda \equiv
\mathbb{Z}[t^{\pm 1},z^{\pm 1}]$.
\end{itemize}
The relevance of such invariants is statistical mechanics and in 
quantum field theory is widely discussed in \cite{Wu,GoRuSi}.

\vspace{12pt}

As will be discussed in the following section, the problem of evaluating,
 or approximating, the Jones polynomial on a quantum computing machine 
has called many people's attention in the last few months. In this respect,
the purely algebraic approach seems somehow lacking in selecting
{\em unitary} representation of the braid group $\mathbf{B}_n$
in a natural way. For instance, the approaches proposed in \cite{AhJoLa}
and  \cite{WoYa} provide two different types of Hilbert space structures 
and associated unitary representations of the braid group by resorting 
to clever, but `ad hoc' constructions.\\
The approach proposed by the authors of \cite{GaMaRa1,GaMaRa2} relies 
on the physical background provided by `unitary' topological quantum field theories,
so that the problem of selecting a suitable unitary representation simply does not exist.
On the other hand, a number of different mathematical tools 
can be called into play (all of which  giving rise to the same
link invariants) but unfortunately no one of them looks simpler than
the purely algebraic method addressed above. In the rest of
this section we shall provide a plain presentation of the quantum group approach
leaving aside many technical details which can be found in
\cite{KiMe}, \cite{MoSt}.
 
\vspace{12pt}

Note preliminarly that, since we are looking for unitary representations (matrices)
to be associated with link invariants, the running variable of the polynomials
(including in particular the Jones polynomial) has to be a unitary complex number
$c \in \mathbb{C}$, with $|c|=1$. The commonly adopted  variable is
a complex, $r$--th {\em root of unity}, namely
\begin{equation}\label{root}
q\,\doteq\, \exp(2\pi\,i/r)\;\;,r \in \mathbb{N},\;r \geq 1
\end{equation} 
and the idea is that, by letting $r$ grow, the polynomial can be evaluated
in more and more points lying on the unit circle in $\mathbb{C}$. The upgraded 
notation for the Jones polynomial is 
\begin{equation}\label{Jones}
\mathsf{J}(L;q)\;\in\;
\mathbb{Z}[q,q^{-1}].
\end{equation}
The invariant of an oriented link $L$ we are going to address is an 
extension of the Jones 
polynomial (\ref{Jones}), denoted by
\begin{equation}\label{copoly1}
\mathsf{J}(L;q;j_1,j_2,\ldots, j_M),
\end{equation}
depending on the  root
of unity $q$ introduced in (\ref{root}) and parametrized by 
labels $\{j_1,j_2,\ldots, j_M\}$ (the `colors') to be assigned to each of 
the $M$ link components $\{L_i\}_{i=1,2,\dots,M}$.
From the point of view of equivalence of links, 
$\mathsf{J}(L;q;j_1,j_2,\ldots, j_M)$ turns out to be 
a `regular isotopy' invariant, but
it can be shown that the quantity
\begin{equation}\label{copoly2}
\frac{q^{-3w(L)/4}}{q^{1/2}-q^{-1/2}}\;\;\mathsf{J}(L;q;j_1,j_2,\ldots, j_M),
\end{equation}
where $w(L)$ is the writhe of the link $L$ defined in
(\ref{writhe}), is invariant under any ambient isotopy transformation.

The {\em colored polynomials} (\ref{copoly2}) reduce to Jones' (\ref{Jones}) when all the colors 
$j_1,j_2,\ldots, j_M$ are equal to a same $j$, with $j=1/2$, but are 
genuine generalizations  as far as they can distinguish knots with the same Jones 
polynomial \cite{Kaul1}. Even more crucially, these invariants are `universal', 
in the sense that they arise from a number of historically distinct
approaches, ranging from R--matrix representations
obtained with the quantum group method, monodromy representations of the braid group
in $2D$ conformal field theories and the quasi tensor category approach by Drinfeld 
up to $3D$  quantum Chern--Simons theory. All these models share the common feature 
of being `integrable', and integrability is reflected by the 
presence of Yang--Baxter--like equations, encoding the algebraic structure 
of braid groups in disguise.

\begin{figure}[htbp]
\begin{center}
\includegraphics[height=6cm]{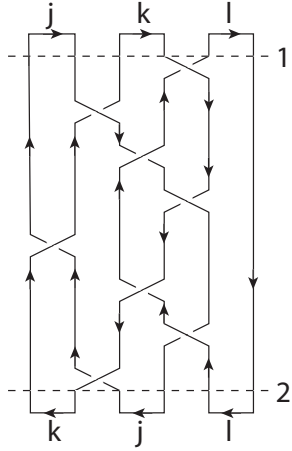}
\end{center}
\caption{A plat presentation of the oriented and colored Borromean link as a closed braid
on six strands.
 The parallel straight lines $1$ ($\equiv \lambda_1$) and $2$ ($\equiv \lambda_2$)
intersect the diagram in points to be associated with Hilbert spaces which inherit the `coloring'
from the corresponding strands.}
\label{borPlat}
\end{figure}

The basic objects which enters into the definition of the invariants (\ref{copoly1}) are
oriented and colored links and braids. Recall that a link is oriented if all its components
$\{L_i\}_{i=1,2,\ldots ,M}$ are endowed with an orientation.
Since $L$ can be thought of as the closure $\hat{\beta}$ of an open braid $\beta$ $\in \mathbf{B}_n$
for some $n$ (Alexander's theorem of section 1) each strand of $\beta$ ($\hat{\beta}$)
inherits naturally an orientation,
depicted in figures by an arrow.\\
The assignment of `colorings' can be carried out in two different ways, by assigning a color either to each
oriented link component $L_m$ $(m=1,2,\ldots,M)$ or to each strand $\mathfrak{l}_i$ $(i=1,2,\ldots,n)$ of the 
associated oriented braid $\beta$ $\in \mathbf{B}_n$. Of course the braid index $n$ is (much) greater
than the number of link components $M$, as can be easily recognized from the samples given 
in Fig. \ref{closures} and Fig. \ref{borPlat}. 
In our paper \cite{GaMaRa2} we adopted the latter option (see also \cite{Kaul2})
 that gives rise to  unitary, $n$--dimensional
representations of $\mathbf{B}_n$ inside the representation ring of the quantum group $SU(2)_q$ defined below.
Here we are going to illustrate the first choice which is technically simpler and
 stresses the role of the so--called $R$--matrix,
namely the set of representations of the crossings in link diagrams as `braiding operators' acting over 
the representation ring of $SU(2)_q$. Note however that both the choices of the colorings provide the same colored link
invariants, possibly up to an overall normalization factor.

Coming at last to the point, the {\em  representation ring of} $SU(2)_q$, 
denoted by $\mathfrak{R}\,(SU(2)_{\,q})$,
can be introduced following in the footsteps of the construction of $SU(2)$--representation theory.
According to our previous notation, the ground ring (in which the link invariants will take their values) is
$\Lambda = \mathbb{Z}[q^{\pm 1}]$ $\subset \mathbb{C}$, with $q=\exp (2\pi i/r)$ as in (\ref{root}).
The elements of $\mathfrak{R}\,(SU(2)_q)$ are complex Hilbert spaces, invariant under the action of the group
(recall that a vector space $V$ is invariant under the action of a group $G$ if $G\times V \rightarrow V$,
namely transformed vectors keep on belonging to $V$; such spaces are referred to as invariant $G$--modules).
As happens for $SU(2)$, it can be shown that $\mathfrak{R}\,(SU(2)_q)$, is spanned by finite--dimensional
 $SU(2)_q$--modules $\{V^j\}$. In the case of $SU(2)$ the labels $\{j\}$ (the spin quantum numbers
from the quantum mechanical point of view) run over all integers and half--integers $\{0,\tfrac{1}{2},1,
\tfrac{3}{2},\ldots\}$, each $V^j$ is characterized by its dimension $(2j+1)$ and is irreducible
(namely cannot be decomposed into a direct sum of invariant subspaces of lower dimensions).\\
 In the $q$--deformed case it can be shown that the $SU(2)_q$--modules $\{V^j\}$ are irreducible if and only if
the labels $\{j\}$ run over the finite set $\{0, \tfrac{1}{2},1,\tfrac{3}{2},\ldots,r\}$. Each $V^j$,
spanned by $(2j+1)$ vectors, can be characterized by a scalar $\in \Lambda$, the $q$--integer
$[2j+1]_q$, where $[\mathfrak{n}]_q\,=$ $(q^{\mathfrak{n}/2}-q^{-\mathfrak{n}/2})/(q^{1/2}-q^{1/2})$
for $\mathfrak{n} \in \mathbb{N}^{+}$, a positive integer. 
Thus, for each choice of the integer $r$, we have a distingushed family of
irreducible representations (irreps) of $SU(2)_q$
\begin{equation}\label{irreps}
\mathfrak{F}_r\,=\,\{\,V^j\,\}_{j=0,\ldots, r}\;\;;\;\;V^j\,\leftrightarrow [2j+1]_q
\end{equation}
which makes $\mathfrak{R}\,(SU(2)_q)$ a finitely generated ring. 
As in the case of $SU(2)$, the ring structure is explicitated 
in terms of the direct sum $\oplus$ and tensor product $\otimes$ of irreps, namely
$$V^j\,\oplus\,V^k\,\in \mathfrak{R}\,(SU(2)_q)\;\text{if}\,j,k\leq r$$
\begin{equation}\label{ring}
V^j\,\otimes\,V^k\,\in \mathfrak{R}\,(SU(2)_q)\;\text{if}\,j+k\leq r, 
\end{equation}
where the ranges of the labels have to be suitably restricted with respect to the standard case.
The analog of the Clebsch--Gordan series, giving the decomposition of the tensor product of
two irreps into a (truncated) direct sum of irreps, reads
\begin{equation}\label{CGser}
V^{j_1}\,\otimes V^{j_2}\;=\;\bigoplus_{j=|j_1-j_2|}^{\text{min} \{j_1+j_2,r-j_1-j_2\}}\;V^j.
\end{equation}
Note however that the ring $\mathfrak{R}(SU(2)_q)$ is much  richer than its `classical' 
$SU(2)$--counterpart because $SU(2)_q$ can be endowed with a {\em 
quasitriangular Hopf algebra} structure. This means that, besides the standard operators
$\oplus$ and $\otimes$ we can also introduce a comultiplication
$\Delta : SU(2)_q \rightarrow$ $SU(2)_q \otimes SU(2)_q$, an antipode map
$A: SU(2)_q \rightarrow SU(2)_q$, a counit $\varepsilon: SU(2)_q \rightarrow
\mathbb{C}$ and  a distinguished invertible element
\begin{equation}\label{Rmatrix}
\mathsf{R}\;\in \;SU(2)_q \otimes SU(2)_q,
\end{equation}                 
called the $R$--{\em matrix}.  We do not insist any further on the explicit
definitions of $\Delta$, $A$ and $\varepsilon$, and  refer to \cite{KiMe} (section 1)
for a quite readable account.\\
The far--reaching role played by the $R$--matrix becomes manifest when we 
define its action on the tensor product of a pair of irreducible 
 $SU(2)_q$--modules in  $\mathfrak{R}(SU(2)_{\,q})$. Denoting by
$\hat{\mathsf{R}}$  the operator associated to $\mathsf{R}$, we have          
\begin{equation}\label{Roper1}                 
\hat{\mathsf{R}}\;:\;\;V^j\, \otimes \,V^k\;\longrightarrow\;
V^k\, \otimes \,V^j,               
\end{equation}              
where, according to (\ref{ring}) the values of the labels $j,k$
have to be suitably restricted. These   
$\hat{\mathsf{R}}$--operators   will be referred to as {\em braiding operators}
associated with the $R$--matrix (\ref{Rmatrix}).             
 If we further extend the action of  $\hat{\mathsf{R}}$ to the ordered product 
of three irreps  $V^j\, \otimes \,V^k \, \otimes V^l$ by defining             
$$\hat{\mathsf{R}}_{jk}\, \doteq \, \hat{\mathsf{R}}\otimes\,\text{Id}:\;
(V^j\, \otimes \,V^k) \, \otimes V^l \,\longrightarrow \, 
(V^k\, \otimes \,V^j) \, \otimes V^l              
$$
\begin{equation}\label{Roper2} 
\; \hat{\mathsf{R}}_{kl}\, \doteq \, \text{Id} \otimes \hat{\mathsf{R}}:\;
V^j\, \otimes \,(V^k \, \otimes V^l) \,\longrightarrow \, 
V^j\, \otimes \,(V^l \, \otimes V^k)\;,  
\end{equation}                
where Id is the identity operator on the corresponding factor, then it can be shown that
these operators satisfy the {\em quantum} Yang--Baxter equation                 
\begin{equation}\label{quantumYB} 
\hat{\mathsf{R}}_{jk} \,\hat{\mathsf{R}}_{kl}\,\hat{\mathsf{R}}_{jk}\,=\,  
\hat{\mathsf{R}}_{kl} \,\hat{\mathsf{R}}_{jk}\,\hat{\mathsf{R}}_{kl}\;.             
\end{equation}                
The adjective `quantum' refers of course to the underlying quantum group setting,
while it easily recognized that                
(\ref{quantumYB}) coincides with
(\ref{algYB2}), the algebraic Yang--Baxter relation characterizing 
the braid group structure, if we perform the substitutions
\begin{equation}\label{compYB}               
\begin{array}{lcl}
\text{ordered triple} \,(j\,k\,l) & \longmapsto & \text{ordered triple} \,(1\,2\,3)\\
\hat{\mathsf{R}}\,(\text{braiding operator}) & \longmapsto & \mathsf{R} \,(\text{crossing})\,.
\end{array}
\end{equation}                  
The explicit expression of the braiding operator $\hat{\mathsf{R}}$ 
(and of its inverse $\hat{\mathsf{R}}^{\,-1}$) can be worked out explicitly
by selecting orthonormal  basis sets in the $SU(2)_{q\,}$--modules 
$V^j, V^k$, for each admissible choice of the pair $j,k$. 
In such bases,
all the braiding operators (\ref{Roper1}) and (\ref{Roper2})
are {\em unitary}. 

Having collected all the necessary algebraic ingredients, the colored invariant
(\ref{copoly1}) for an oriented link $L$ with $M$ components can be now consistently interpreted 
as a single, $\Lambda$--linear map
\begin{equation}\label{copoly3}              
\mathsf{J}\,(L;q;j_1,j_2,\ldots, j_M)\,:\,\;\;
\mathfrak{R}\,(SU(2)_q)\;\longrightarrow \;\Lambda\, ,
\end{equation}
where the choice of the integer $r$ in the root of unity 
(\ref{root}) is constrained by the requirement
$r\geq M$, at least in the most general case ($M$ distinct colors).

The prescription for working out $\mathsf{J}\,(L;q;j_1,j_2,\ldots, j_M)$
can be summarized as follows (compare also Fig. \ref{borPlat}, where
the plat presentation of the Borromean link is depicted).

\begin{itemize}
\item Present the link $L\,=\,\cup\, L_i \;(i=1,2,\ldots,M)$ as the plat closure
of a braid and choose an orientation for each component (depicted by an arrow).
Assign to each component a (distinct) color
\begin{equation}\label{colL}
L_i\;\;\;\longrightarrow\;j_i\;\;\;\;(i=1,2,\ldots,M)\,.
\end{equation}
\item Insert two parallel horizontal lines $\lambda_1$,
$\lambda_2$ cutting the `cap' and `cup' 
portions of the diagram, respectively. This choice
provide the diagram with an overall, downward
orientation.\\
The region of the diagram lying between 
$\lambda_1$ and $\lambda_2$ is an open braid whose strands inherit
suitable labels from the colorings (\ref{colL}).
\item Assign to the intersection point between a line ($\lambda_1$ or $\lambda_2$)
and the string labelled by $j$ the $SU(2)_q$ irreducible module
$V^j$ belonging to the distinguished family defined in (\ref{irreps}).\\
The whole configurations of intersection points on $\lambda_1$
and $\lambda_2$, each ordered from left to right, are to be associated with 
the $SU(2)_{q\,}$--modules $V_{\lambda_1}$ and $V_{\lambda_2\,}$, respectively,
where each of them is the ordered tensor product of the individual irreps.
To give an explicit expression of these correspondences, consider the case 
of the $3$--components Borromean link depicted in 
Fig. \ref{borPlat}, where in particular 
$$V_{\,\lambda_1}\,=\,V^j \otimes V^j \otimes V^k \otimes V^k \otimes V^l
\otimes V^l
$$
\begin{equation}\label{Vline}
V_{\,\lambda_2}\,=\,
V^k \otimes V^k \otimes V^j \otimes V^j \otimes V^l
\otimes V^l.
\end{equation}
Note that $V_{\,\lambda_1}$ and $V_{\,\lambda_2}$ have the same dimension
as Hilbert spaces over $\mathbb{C}$, given by the product of the
dimensions of the individual factors. The number  of such factors,
say $2N$, is the same for the two spaces and equals the number of
strands of the braid, or even the number of
`caps' (`cups') lying above the line $\lambda_1$ (below
$\lambda_2$) divided by two. This feature derives of course  from
the topological presentation  we adopted for the link $L$, since  
the braid obtained from the plat closure of any link  has an
even number of strands.
\item Going on with the example, in the representation ring
$\mathfrak{R} (SU(2)_{q\,})$ there exists a well defined, unitary operator
$\hat{\mathsf{B}}(L;q;j,k,l)$ to be associated with the trasformation
 relating $V_{\,\lambda_1}$ and $V_{\,\lambda_2}$ in 
the diagram of the Borromean link $L$ in Fig. \ref{borPlat}
\begin{equation}\label{Borroper}
\hat{\mathsf{B}}(L;q;j,k,l):
V_{\,\lambda_1} \;\longrightarrow
V_{\,\lambda_2},
\end{equation}
where $V_{\,\lambda_1}$ and $V_{\,\lambda_2}$ are explicitly
defined in (\ref{Vline}).
The composite braiding operator 
$\hat{\mathsf{B}}(L;q;j,k,l)$ can be decomposed into an ordered 
sequence of the `elementary' unitary braiding operators 
$\hat{\mathsf{R}}$ (and their inverses) introduced in (\ref{Roper1}), 
suitably tensorized with identities. 
The sequence is uniquely determined by going trough 
the diagram from $\lambda_1$ to $\lambda_2$.
\item In the case of the Borromean link, the matrix elements
of the braiding operator (\ref{Borroper}) evaluated on
(the tensor product of) orthonormal basis vectors
of the spaces $V^j$,$V^k$, $V^l$ can be collected into a unitary
$(2J+1) \times (2J+1)$ matrix parametrized by the colors $j,k,l$, namely 
\begin{equation}\label{Borrmatrix}
 \mathsf{B}_{\alpha \beta}\,(j,k,l) \;\in\; U(\Lambda,\,2J+1)\;\; (\alpha, \beta
 =1,2,\ldots, 2J+1),
\end{equation}
where $U(\Lambda,\,2J+1)$ is the  algebra of unitary matrices on
the ground ring $\Lambda \equiv \mathbb{Z}[q,q^{-1}]$
and $(2J+1)=(2j+1)(2k+1)(2l+1)$.\\
Finally, the colored link invariant $\mathsf{J} (L;q;j,k,l)$
is obtained by taking the trace of the matrix (\ref{Borrmatrix}),
formally 
\begin{equation}\label{Borrtrace}
\mathsf{J} (L;q;j,k,l)\,=\,(\text{Tr}\,\mathsf{B}_{\alpha \beta} )\,(j,k,l),
\end{equation}
where the resulting quantity turns out to contain the colorings through
the `quantum weights'
$[2j+1]_q$, $[2k+1]_q$, $[2l+1]_q$ (refer to the appendix of \cite{Kaul2}
for the explicit expression of (\ref{Borrtrace})).
\end{itemize}

The above list of prescriptions can be applied to any link diagram as well,
the `output' being the colored link polynomial of the associated link,
namely the (colored) trace of a suitable unitary matrix.
Such `Markov traces' are exactly the objects that can be handled in a
quantum computational framework, as will be illustrated in the next section.
In the present context the original Jones polynomial (\ref{Jones}) can be easily recovered
by choosing the fundamental ($\tfrac{1}{2}$--spin) $2$--dimensional 
irreducible representation on each of the link components $\{L_i\}_{i=1,\ldots,M}$. 

Note in conclusion that the whole construction   does not depend on
the particular plat presentation we choose for the link under examination, since
the quantum Yang--Baxter equation (\ref{quantumYB}) ensures that
the braiding operators associated with different plat presentations
of a same link can be converted one into the other. In other words, the
role of this operatorial identity is specular 
to the role played by  Markov move of
type {\bf i)} at the purely topological level. As indicated in
(\ref{copoly2}),
the complete (ambient isotopy) invariance of the colored polynomials,
implemented topologically by both Markov moves  {\bf i)} and {\bf  ii)},
can be restored by taking into account the writhe
of the link.

\section{Quantum computation of link polynomials}

Having defined in the second part of the previous section the $SU(2)_q$--colored link polynomials,
let us focus now on the Jones invariant (\ref{Jones}), which is the simplest, 
$\tfrac{1}{2}$--spin colored polynomial, on the one hand, and the prototype of
invariants arising in a purely algebraic context, on the other ({\em cfr.} the
discussion at the beginning of section 3). The reason why Jones' case is so
crucial in the computational context is actually due to the fact that a `simpler'
link invariant, the Alexander--Conway polynomial, can be computed efficiently, while
the problem of computing $2$--variable polynomials --such as the HOMFLY invariant
briefly addressed in section 3-- is $\mathbf{NP}$--hard (see \cite{BiBr} for the definitions
of the mentioned invariants and
\cite{JaVeWe} for an account of computational questions).
The issue of computational complexity of the Jones polynomial in classical information theory
can be summarized as follows.
\begin{quote}
{\em \underline{Problem 3} How hard is it to determine the Jones polynomial\\
 of a link $L$? }
\end{quote} 
A quite exhaustive answer has been provided in \cite{JaVeWe}, where the evaluation of the Jones polynomial
of an alternating link $\tilde{L}$ at a root of unity $q$ is shown to be 
$\mathbf{\# P}$--hard, namely computationally intractable in a very strong sense (see the
definition of this class below).\\ 
A number of remarks are in order. Recall first that
`alternating' links are special instances of links, the planar diagrams of which exhibit
over and under crossings, alternatively. Thus, the evaluation of the invariant of generic,
non--alternating links is at least as hard. Secondly, the computation becomes feasible 
when the argument $q$ of the polynomial is a  2nd, 3rd, 4th, 6th  root of unity 
(refer to \cite{JaVeWe} for details on this technical issue). 
Recall finally  that the $\mathbf{\# P}$ complexity class can be defined as the class of enumeration problems
in which the structures that must be counted are recognizable in polynomial time.
A problem $\pi$ in $\mathbf{\# P}$ is said $\mathbf{\# P}$--complete if, for any other problem $\pi '$ in
$\mathbf{\# P}$, $\pi '$ is polynomial--time reducible to $\pi$; if a polynomial time algorithm were found
for any such problem, it would follow that $\mathbf{\# P} \subseteq \mathbf{P}$. A problem is $\mathbf{\# P}$--hard
if some $\mathbf{\# P}$--complete problem is polynomial--time reducible to it. Instances of 
$\mathbf{\# P}$--complete problems are the counting of Hamiltonian paths in a graph and the most intractable
problems arising in statistical mechanics, such as the enumeration of configurations contributing to  ground 
state partition functions. 

The intractability of Problem 3 relies on the fact that it is not possible to recognize in polynomial time
all the equivalent configurations of a same link $L$, namely link diagrams $\{D(L), D'(L), D''(L),\ldots \}$
related to each other by (regular) isotopy. Coming back to the problems addressed in section 2, since any
link can be presented efficiently as a closed braid (Problem 1A), we are justified in switching
our attention to closed braids and implementing regular isotopy of diagrams by means of Markov move
of type {\bf i)}. However, the intractability of Problem 1B (selecting the diagram with the minimum braid
index) prevents us from selecting an `optimal' representation of the isotopy class of diagrams that would
provide, in turn, a unique standard configuration to be handled for computational purposes. Moreover, owing to the fact that Markov
move {\bf i)}  is closely related to the conjugacy problem in the braid group (Problem 2B), the whole matter
could be reformulated within the framework of the theory of finitely presented groups as well. Then the issue of
the optimal presentation turns out to be related also with the $\mathbf{NP}$--complete problem stated in 2A' of section 2
(the `shorter word' problem).

In the discussion above, the relevant quantities encoding the `size' of a typical instance of the computational
problem --a link diagram $L$ presented as a closed braid on $n$ strands, $L=\hat{\beta}$ as in (\ref{beta})-- 
are of course the number
of crossings $\kappa$ and the braid index $n$. We might consider, instead of $\kappa$, the length $|\beta |$ of the open
braid $\beta$ associated with $L$, which equals the number of generators (and inverse generators) in the explicit
expression of $\beta$ as a word in $\mathbf{B}_n$. Finally, also the argument $q$ of the Jones polynomial (\ref{Jones})
is a relevant parameter since, when the integer $r$ in (\ref{root}) becomes $\gg 1$, 
we would reach more and more points on the unit circle in the complex plane, thus giving more and more accurate
evaluations of the invariant.

\vspace{12pt}

The computational intractability of Problem 3 does not rules out by any means the possibility of `approximating'
efficiently Jones invariant. 
\begin{quote}
{\em \underline{Problem 4}. How hard is it to approximate the Jones polynomial $\mathsf{J}(L,q)$ of a link
$L$ at a fixed root of unity $q$ ($q \neq$ 2nd, 3rd, 4th, 6th root)?}
\end{quote}
Loosely speaking, the approximation we are speaking about is a number $Z$ such that, for any choice
of a small $\delta > 0$, the numerical value of $\mathsf{J}(L,q)$, when we
substitue in its expression the given value of $q$,  differs from $Z$ by an amount
ranging  between $-\delta$ and $+\delta$. In a probabilistic setting (either classical or quantum)
we require that the value $Z$ can be  accepted as an approximation of the polynomial if 
\begin{equation}\label{apprJ}
\text{Prob}\;\left\{|\,\mathsf{J}(L=\hat{\beta},q)\,-\,Z\,|\;\leq \delta \right\}\,\geq\,\frac{3}{4},
\end{equation}
where we agree to present always the link as a closed braid and refer the reader to
\cite{AhJoLa}  and \cite{WoYa} for more accurate statements of (\ref{apprJ}).\\
In the framework of classical complexity theory there do not exists algorithms to handle Problem 4
and thus, at least at the time being, this problem is to be considered as intractable.
From the quantum computational side, it was Michael Freedman who first addressed the general 
problem of evaluating quantities of topological nature arising in the context of
$3D$--topological quantum field theories and associated $2D$--conformal field theories \cite{Fre}
(the physical quantum systems to be simulated are typically anyonic systems).
Freedman's `quantum field computer' was especially designed to this goal, although it was later recognized
that this model of computation is actually (polynomial time) reducible to the standard quantum circuit model.
The formal statement of the answer to Problem 4 was given in \cite{BoFrLo} (see also \cite{FrLaWa}) 
and can be summarized as follows.
\begin{quote}
{\em 4.} The approximation of the Jones polynomial of a link presented as the closure of a braid
at any fixed root of unity is $\mathbf{BPQ}$--complete. Moreover, this problem is universal
for quantum computation, namely is the `prototype' of all
problems efficiently solvable on a quantum computer.
\end{quote}
Recall that
$\mathbf{BQP}$ is the computational complexity class of 
problems which can be solved in polynomial time by a quantum computer 
with a probability of success at
least $\frac{1}{2}$ for some fixed (bounded) error. In \cite{BoFrLo} it was
proved that $\mathbf{P}^{\mathsf{J}}$ $=\mathbf{BQP}$, where $\mathbf{P}^{\mathsf{J}}$ is defined as the 
class of languages accepted
in polynomial time by a quantum Turing
machine with an oracle for the language defined by  Problem 4. This equality between 
computational  classes implies that, if we find out  an efficient quantum algorithm
for Problem 4, then the problem itself is complete for the class $\mathbf{BQP}$, 
namely  each problem in this last class can be efficiently reduced to 
a proper approximate evaluation of the Jones polynomial of a link
(see \cite{WoYa} for a detailed discussion on this issue). According to the above remarks, 
the goal of working out efficient quantum algorithms for Problem 4 (not explicitly
given in \cite{BoFrLo}) does represent a breakthrough in quantum information theory.

\vspace{12pt}

In the rest of this section we are going to illustrate,
without entering in much technical details, the efficient quantum
algorithms recently proposed by three groups, Aharonov, Jones and Landau \cite{AhJoLa},
Garnerone, Marzuoli and Rasetti \cite{GaMaRa1,GaMaRa2}, Wocjan and Yard \cite{WoYa}.\\
Generally speaking, the approaches proposed in \cite{AhJoLa,WoYa} and in \cite{GaMaRa1,GaMaRa2}
differ  both in the theoretical background underlying the construction of link invariants,
and in the model of quantum computation used to deal with calculations.

The Jones polynomial is defined in \cite{AhJoLa,WoYa} by resorting to the purely algebraic 
framework outlined in the first part of section 3, and these authors focus on the search
for a {\em unitary} representation of the braid group $\mathbf{B}_n$ in the Temperley--Lieb
algebra \cite{AhJoLa} or in the Hecke algebra \cite{WoYa}. However, as already pointed out in section 3, 
such representations are selected on the basis of purely formal criteria, and the deep connection
between Jones polynomial and topological quantum field theory \cite{Wit} is left aside.\\
In our approach the universality of Jones polynomial (or, better, of the colored link
invariants (\ref{copoly1})) in so many physical contexts makes it possible, not only to select
a `natural' unitary representation of the braid group, but justifies also the use of anyone
of the {\em equivalent} mathematical backgrounds. In this paper we have described the
colored polynomials in the algebraic framework provided by the representation ring
$\mathfrak{R}(SU(2)_{\,q})$ (and associated R--matrix) since it does not require any
previous knowledge of quantum field theoretic notions, not so familiar to mathematicians
and computer scientists, either to experimental physicists.
 Fully equivalent methods based on $3D$--Chern--Simons theory
and associated Wess--Zumino--Witten $2D$--conformal field theory could have been used as well
\cite{GaMaRa1,GaMaRa2}, making our approach closer to the spirit of Freedman's original paper \cite{Fre}.
Moreover, the $SU(2)_{q\,}$--colored link invariants are more general than the Jones
polynomial with respect to the detection of knots \cite{Kaul1}, on the one hand, and can be related 
in a quite simple way to more general topological invariants characterizing closed 
three--dimensional spaces, on the other \cite{Lic,KiMe,MoSt}. The latter remark opens the
intriguing possibility of addressing, on an effective computational ground,
basic questions in field theories whose dynamical variables are geometric objects,
including quantum gravity models.

For what concerns the computational frameworks, while the papers \cite{AhJoLa,WoYa} rely
on techniques based on the standard quantum circuit model, in \cite{GaMaRa1,GaMaRa2}
we exploit the spin--network quantum simulator proposed in \cite{MaRa} to set up
 quantum--automaton--like implementations of braiding operators, each 
elementary step being worked in one unit of the intrinsic discrete  time
of the automaton itself. We are currently completing the proof that 
each of the latter elementary unitary operations can be `efficiently' implemented 
also with respect to the `standard' quantum circuit model \cite{GaMaRa3}.

The results of \cite{AhJoLa} can be summarized as follows (see also the recent review
 \cite{KaLo}). The unitary representation 
of the braid group $\mathbf{B}_n$ in the Temperley--Lieb algebra 
({\em cfr.} the first part of section 1 above for the algebraic background) 
is  obtained by resorting to an adaptation of the 
`path model' proposed earlier by  Jones himself. 
The key idea is that each generator of the
braid group is mapped by the path model representation to a unitary matrix which can 
be simulated efficiently by a quantum circuit. In particular, each crossing can be implemented by
a quantum circuit using polynomially many elementary quantum gates, the resulting unitary
matrix being no longer `local' (namely, it operates non trivially on
more than just two qubits corresponding to the two strands concurring at the crossing point), 
but in any case efficiently implemented.
An entire braid (whose plat closure represents a link $L$) can be applied efficiently
by employing \underline{a number of elementary gates} \underline{polynomial in the braid index $n$} and
\underline{in the number of crossings $\kappa$}. 
The approximation the Jones polynomial is reduced to the approximation of the Markov trace
of the unitary matrix associated with the braid, which is carried out by a
standard quantum algorithmic technique (the Hadamard test).
In the work \cite{WoYa} more general closures of braids (not just plat closures) are addressed, 
and the algorithm is based on a local qubit implementation of the unitary Jones--Wenzl 
representations of the braid group. \\
In both the latter papers,  the value of the integer $r$ in the root of unity $q$
is arbitrarily chosen, but constant, while a polynomial estimate of the growth
of the time complexity function with respect to this parameter too has been
established in \cite{GaMaRa2} by resorting to field theoretic arguments.

As already pointed out, our  approach is quite different from the previous ones
in working out  the unitary matrix to be associated with the plat closure of a braid,
while the approximation of its trace, giving the colored link polynomials as in the example
(\ref{Borrtrace}), can be carried out by similar standard techniques \cite{GaMaRa3}.
To enter in some more details, the `physical' background provided by the 3$D$ quantum $SU(2)$
Chern--Simons field theory  plays here a prominent role,
because the computational scheme of the spin--network simulator \cite{MaRa} is actually designed
as a discretized conterpart of  the topological quantum
computation setting proposed in \cite{Fre,FrLaWa}. Our framework is designed to deal
with the $SU(2)_q$--colored link polynomials (\ref{copoly3}) viewed
as vacuum expectation values of composite 
`Wilson loop' operators in Chern--Simons theory, on the one 
hand, and with unitary representations
of the braid group, on the other.
These expectation values, in turn, provide a bridge between the theory of formal languages 
and quantum computation, once more having as natural  arena for discussion the 
spin--network environment.
 We actually implement families of
finite states (and discrete time)--quantum automata 
capable of accepting the language generated by the braid group,
and whose transition amplitudes are colored polynomials.
In other words, our results can be interpreted in terms of 
`processing of words' --written in the alphabet given by the generators of the braid 
group in the given representation-- on a quantum automaton in such a way 
that the expectation value associated with the internal automaton
`evolution' is exactly 
the required link polynomial (after the application of a suitable `trace operation'
and the approximation of the trace within some fixed range as required in (\ref{apprJ})). 
Our quantum automaton calculation provides the unitary matrix associated with
(the plat closure of) a colored link $L$ on $n$ strands in a number of steps which
is bounded \underline{linearly in the number of
crossings $\kappa$ of the link}, on the one hand, and \underline{polynomially
bounded in terms of the braid index $n$} ($\leq n\ln n$), on the other.


\begin{thebibliography}{99}

\bibitem{Wit}
E. Witten, Commun. Math. Phys. {\bf 121}, 351 (1989).

\bibitem{Jo85}
V.F.R. Jones, Bull. Amer. Math. Soc. {\bf 12}, 103 (1985).

\bibitem{NiCh} M.A. Nielsen, I.L. Chuang,
\textit{Quantum Computation and Quantum Information}
(Cambridge University Press, 2000).

\bibitem{Wu}
F.Y. Wu, Rev. Mod. Phys. {\bf 64}, 1099 (1992).

\bibitem{GoRuSi}
C. Gomez, M. Ruiz--Altaba, G. Sierra,
\textit{Quantum Group in Two--dimensional Physics}
(Cambridge University Press, 1996).

\bibitem{AhJoLa}
D. Aharonov, V.F.R. Jones, Z. Landau,
\textit{A polynomial quantum algorithm for approximating the Jones polynomial},
arXiv: quant-ph/0511096 (2005).

\bibitem{BoFrLo} 
M. Bordewich, M. Freedman, L. Lovasz, D. Welsh, 
\textit{Approximate counting and quantum computation}, 
to appear in Combinatorics, Probability and Computing (2006).

\bibitem{GaMaRa1}
S. Garnerone, A. Marzuoli, M. Rasetti,
J. Phys.: Conf. Ser. {\bf 33}, 95 (2006).

\bibitem{GaMaRa2}
S. Garnerone, A. Marzuoli, M. Rasetti,
\textit{Quantum automata, braid group and link polynomials},
arXiv: quant-ph/0601169 (2006).


\bibitem{WoYa}
P. Wocjan, J. Yard, 
\textit{The Jones polynomial: quantum algorithms and applications
to quantum complexity theory},
arXiv: quant-ph/0603069 (2006).

\bibitem{Rol}
D. Rolfsen, \textit{Knots and Links},
(Publish or Perish, Berkeley, CA, 1976).

\bibitem{Bir}
J.S. Birman \textit{Braids, links, and mapping class groups},
(Princeton University Press, 1974)

\bibitem{BiBr}
J.S. Birman, T.E. Brendle,
\textit{Braids: a survey}, 
arXiv: mathGT/0409205 (2004).

\bibitem{Lic}
W.B.R. Lickorish,
\textit{An Introduction to Knot Theory},
(Springer--Verlag, New York, 1997).

\bibitem{MaKaSo}
W. Magnus, A. Karrass, D. Solitar,
\textit{Combinatorial Group theory},
2nd Ed. (Dover Publ., New York, 1976).


\bibitem{AnAnGo}
I. Anshel, M. Anshel, D. Goldfeld,
Math. Res. Lett {\bf 6}, 1 (1999).



\bibitem{HOMFLY}
P. Freyd, D. Yetter, J. Hoste, W. Lickorish, K. Millett, A. Ocneanu,
Bull. Amer. Math. Soc. {\bf 12}, 183 (1985).



\bibitem{Jo05}
V.F.R. Jones, \textit{The Jones polynomial},\\
from http://math.berkeley.edu/vfr/.


\bibitem{KiMe}
R. Kirby, R. Melvin,
Invent. Math. {\bf 105}, 473 (1991).


\bibitem{MoSt}
H.R. Morton, P. Strickland,
Math. Proc. Camb. Phil. Soc. {\bf 109}, 83 (1991).


\bibitem{Kaul1}
P. Ramadevi, T.R. Govindarajan, R.K. Kaul,
Mod. Phys. Lett. A {\bf 9}, 3205 (1994).

\bibitem{Kaul2}
R.K. Kaul, Commun. Math. Phys. {\bf 162}, 289 (1994).

\bibitem{JaVeWe}
F. Jaeger, D.L. Vertigan, D.J.A. Welsh,
Math. Proc. Camb. Phil. Soc. {\bf 108}, 35 (1990).






\bibitem{Fre}
M.H. Freedman, 
Pro. Nat. Acad. of Science USA {\bf 95}, 98 (1998).

\bibitem{FrLaWa} 
M.H. Freedman, M.Larsen, Z. Wang, 
Commun. Math. Phys. 
{\bf 227}, 605 (2002)

\bibitem{MaRa}
A. Marzuoli, M. Rasetti, Ann. Phys. {\bf 318}, 345 (2005).

\bibitem{GaMaRa3}
S. Garnerone, A. Marzuoli, M. Rasetti (in preparation).

\bibitem{KaLo}
S.J. Lomonaco, L.H. Kauffman,
\textit{Topological quantum computing and the Jones polynomial}
 arXiv: quant-ph/0605004 (2006).
 
 
\end{thebibliography}
\end{document}